\documentstyle[12pt]{article}
\input epsf
\begin{document}
\newcommand{\lsi}{\,\raisebox{-0.13cm}{$\stackrel{\textstyle<}
{\textstyle\sim}$}\,}

\rightline{RU-95-82}
\rightline{hep-ph/9512306}

\baselineskip=18pt
\vskip 0.7in
\begin{center}
{\bf \LARGE New Signature of Squarks }\\ 
\vspace*{0.4in}
{\large Glennys R. Farrar}\footnote{Research supported in part by
NSF-PHY-94-23002} \\
\vspace{.1in}
{\it Department of Physics and Astronomy \\ Rutgers University,
Piscataway, NJ 08855, USA}\\
\end{center}
\vspace*{0.2in}
\vskip  0.9in  

{\bf Abstract:} When the gluino is light and long lived, missing
energy is a poor signature for both squarks and gluinos.  Instead,
$S_q S_q^*$ production in $e^+ e^-$ and $p \bar{p}$ collisions
characteristically results in events with $\ge 4$ jets.  Methods are
proposed for deciding whether an observed excess of 4-jet events is
due to $S_q S_q^*$ production.  The recent report by ALEPH of
observation of 14 4-jet events when 7 were expected is discussed. 

\vskip 0.5in
\begin{center}
{\it Dec. 13, 1995; elaborated Dec. 22, 1995}
\end{center}
\thispagestyle{empty}
\newpage
\addtocounter{page}{-1}
\newpage

I have recently outlined\cite{f:99,f:101,f:102,f:103} some of the
low-energy features of theories in which dimension-3 SUSY breaking
operators are highly suppressed.  This is the generic situation in
several interesting methods of SUSY breaking.  Two to four free
parameters of the usual minimal supersymmetric standard model ($A$ and
the gaugino masses) vanish at tree level.  The elimination of these
SUSY breaking operators implies that there is no additional CP
violation at T=0  beyond what is already present in the standard
model\cite{f:101}, avoiding the embarassing SUSY CP problem.  Gauginos
are massless at tree level but get calculable masses through radiative
corrections from electroweak and top-stop loops. Evaluating these
leads to gluino and photino masses less than $\sim 1$ GeV\cite{f:99};
the lightest chargino has a mass $ \le m_W$ and thus should be
discovered at LEP II. 

The lightest $R$-hadron is a gluino-gluon bound state
often called $R^0$.  Its mass should be in the range $\sim 1.2 - 2.2$
GeV\cite{f:95,f:101}.  It decays to a photino and hadrons.  The
photino is an attractive dark matter candidate, having the correct
abundance for $r \equiv m(R^0)/m_{\tilde{\gamma}}$ in the range $\sim
1.6 - 2$\cite{f:100}, compatible with expectations\cite{f:99}.  For
$r$ in this range, $\tau(R^0) \sim (10^{-7} - 10^{-10})
(\frac{M_{sq}}{100 {\rm GeV}})^4$sec, so that the lifetime of the
$R^0$ can be long enough that it would not have been detected in
existing searches\cite{f:95}. Methods to study the $R^0$ and other
$R$-hadrons experimentally are given in \cite{f:101,f:102}.   

Squarks are pair produced in $e^+ e^-$ annihilation with cross section
$\frac{1}{2} \beta^3$ times that of a massless quark of the same
flavor and chirality\footnote{Although squarks have spin-0, they can
be ascribed chirality because supersymmetry associates them with a
quark of definite chirality.  As discussed below, squark chirality
violation is very small in this scenario, except for top-squarks.}. In
a hadron collider they can be produced either in pairs from $q
\bar{q}$ annihilation and gluon-gluon fusion, or singly in association
with a gluino or (at the $\sim 1\%$ level) a photino, in quark-gluon
fusion. At an $e^- p$ collider they are produced singly in association
with a gluino or (at the 2\% level) a photino.  

Once produced, squarks decay dominantly via $S_q \rightarrow q +
\tilde{g}$.  The gluino hadronizes forming a jet, due to the long
lifetime of the $R^0$.  This is to be contrasted with the conventional
case of a very heavy and thus short lived gluino, for which the
photino production is prompt.  When the $R^0$ in the gluino jet 
finally decays, the energy carried by the photino is so small that the
conventional missing energy signature is not useful\footnote{Squarks
decay directly, thus promptly, to a photino and quark with a branching
fraction $Q_{sq}^2 \alpha_{em}/(\frac{4}{3} \alpha_s)$.  For a charge
2/3 squark this occurs about 2\% of the time, the same factor which
appears in the ratio of photino to gluino production in deep inelastic
scattering from an up quark.  Averaging over the $u,~d,~s,~c$ and
$b$ squarks, $1\%$ of the time a single prompt photino is present in
the decay products of a squark pair.  Naively rescaling the UA(1) and
Tevatron collider limits, to account for this loss in sensitivity,
leads to limits much worse than those obtained below from the $Z^0$
hadronic width.}.  Existing Tevatron collider limits do not apply.
The remainder of this paper is devoted to establishing a search
procedure appropriate to this scenario. 

The best limit on squark masses prior to LEP 1.5, if missing energy is
not useful, comes from the determination of the hadronic width of the
$Z^0$.  In $e^+ e^- $ collisions, $\sigma(e^+ e^- \rightarrow
S_q S_q^*) = \frac{1}{2} \beta^3 \sigma(e^+ e^- \rightarrow q
\bar{q})$ neglecting quark mass, for any given flavor and chirality.
Therefore production of a $(u_L,~ u_R,~ d_L,~d_R)$-type squark
antisquark pair would increase the total hadronic width of the $Z^0$
by a fraction (0.06, 0.01, 0.09, 0.003)$\beta^3$.  The limit on
``extra'' hadronic width of the $Z^0$ then limits the mass of squarks.
If there are five degenerate ``light'' squarks (called the dls case
below), their mass must be greater than $\sim M_Z/2$.  If only a
single flavor of squark is light, this limit is greatly reduced.
Considering $e^+ e^- \rightarrow S_q \bar{q} \tilde{g} + S_q^* q
\tilde{g}$ and virtual corrections to $e^+ e^- \rightarrow q \bar{q}$
allows the dls limit to be improved to $50-60$
GeV\cite{clavetalsquarks,bhat_deltaR}.  The analysis should be redone
with new $Z^0$ width values and careful treatment of the value of
$\alpha_s(M_Z)$ and its running to lower scales, assuming a light
gluino, since the expected $Z^0$ width is sensitive to this. 

For squark masses up to $\sim \frac{1}{2} E_{\rm LEP}$, $S_q S_q^*$
pairs can be readily produced and identified at LEP.  At larger
masses, the Tevatron collider complements LEP.  In both cases one
studies events with four or more jets, as discussed below.  QCD
background is much more severe at the hadron collider, but the
signal-to-noise may actually improve with increasing squark mass.  The
signature of squarks appears to be less distinctive at HERA, but the
fact that only $u$ and $d$ squarks are produced, so the prompt
photino fraction is enhanced, means that if squarks are found HERA
can give complementary information to the other machines.  In the
following I will concentrate on $S_q S_q^*$ production at LEP, with
the obvious parallels to the Tevatron and HERA searches left implicit. 

\begin{figure}
\epsfxsize=\hsize
\epsffile{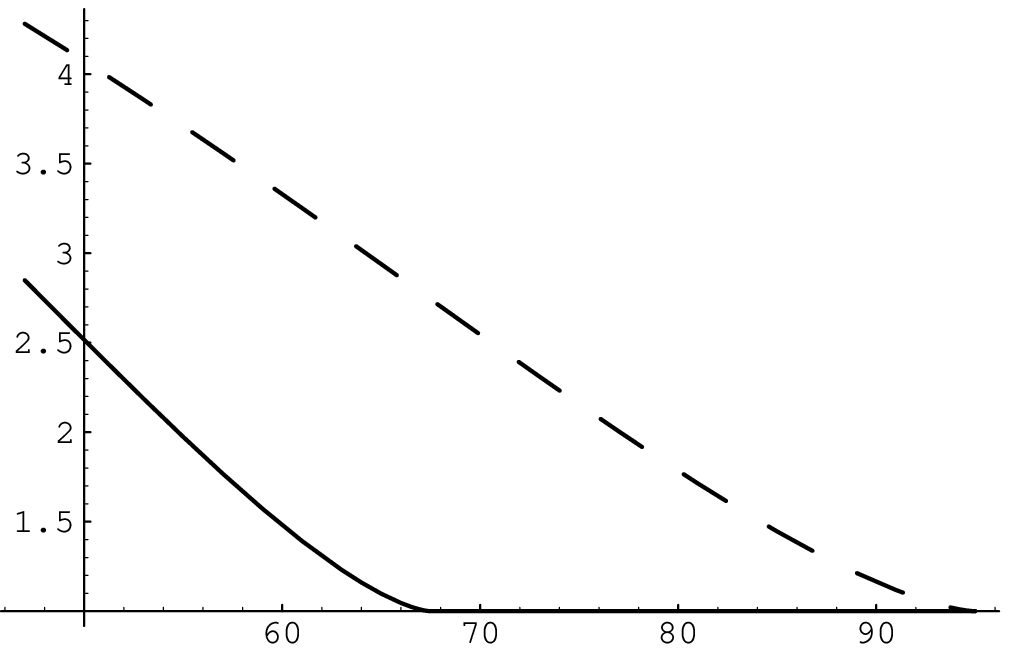}
\caption{Ratio, $R_{\ge4}$, of the number of events with $n_{jet} \ge
4$ in the presence and absence of $u,d,s,c,b$ squarks, as a
function of the common squark mass in GeV.  Solid and dashed curves
are for $E_{cm} = 135$ and 190 GeV, respectively.} 
\label{sq}
\end{figure}

In $e^+ e^-$ collisions at the $Z^0$ and above, about 10\% of hadronic
events are observed to consist of four or more jets, defining jets
with $y_{cut} = 0.01$.  Since essentially every $S_q S_q^*$ pair
produces four  or more jets, a large enhancement of four-jet events is
expected when one is sufficiently above threshold that the $\beta^3$
suppression is not severe.  To be more quantitative, define $f_{\ge4}$
to be the fraction of ordinary events with four or more jets for a
given energy and jet-finding algorithm, and  
\begin{equation}
r_i(m_i,E) \equiv \frac{\sigma(e^+ e^- \rightarrow S_q^i 
S_q^{i*})}{\Sigma_i ~\sigma(e^+ e^- \rightarrow q^i \bar{q}^i)}.
\end{equation}
Then the ratio of the number of $n_{jet} \ge 4$ events with and
without squarks is $R_{\ge4}(E) = \frac{\Sigma_i~ r_i(m_i,E) +
f_{\ge4}}{f_{\ge4}}$.  To show how large an effect squarks produce,
Fig. \ref{sq} plots $R_{\ge 4}(m,E)$ for $E=135$ and 190 GeV, for
degenerate $u,~d,~s,~c,~b$ squark (dls) masses.  It should be stressed
that other constraints such as the $\rho$ parameter and Tevatron top
quark studies exclude a too-light sbottom and stop\footnote{Since the
gluino here is light, unless the stop is approximately degenerate with
the top or heavier, the main decay mode of the top would be $t
\rightarrow S_t + \tilde{g}$ which is excluded by the consistency
between the observed top mass and the rate of its observation in
conventional signatures.  The sbottom mass must be $>~ \sim
\frac{3}{4}$ the stop mass in order not to produce too large a change
in the $\rho$ parameter.}, so the dls case is not realistic.

In the recent LEP 1.5 run at $E_{cm} = 130-140$ GeV,
ALEPH\footnote{L. Rolandi, Joint CERN Particle Physics Seminar on
First Results from LEP 1.5, Dec. 11, 1995} found 14 events which meet
their 4-jet criteria, when 7.1 events are expected from standard model
physics and less than one 4-jet event is expected from either $hA$ or 
$H^+H^-$ production.  With dls, this rate of 4-jet events implies a 55
GeV common squark mass, as can be seen from Fig. 1.  Note that
$R$-squarks decouple at this energy because the photon and $Z^0$
contributions just cancel.  Thus only $L$-squarks are probed at this
particular energy.  Furthermore, at this energy $U$- and $D$-
type squarks are produced equally, if they have the same $\beta^3$
factor, making it easy to rescale from the unrealistic dls case.  

In view of the various constraints which limit the number of light
squarks, and exclude the stop and sbottom being so light, possibly the
most plausible explanation of the ALEPH events, if they are real, would 
be a ``cocktail'' of pair production of a couple of flavors of squarks
with 55 GeV masses and pair production of a slightly heavier chargino.
In the no-dimension-3-SUSY breaking scenario, one chargino must be
lighter than the $W$, so this is a natural possibility if there are
such light squarks.  In this case the dominant chargino decay mode 
would be $\chi^\pm \rightarrow S_q \bar{q} + S_q^* q$, assuming the
sneutrinos are heavier.  If such a decay channel is available, the 
branching ratio of the charginos to the $f \bar{f}
\chi^0$ final state (via a virtual $W$) would be suppressed, so the
chargino would not have shown up in conventional searches.  Although
the ultimate final state of such chargino pairs has $\ge 6$ jets, for
a small mass splitting between chargino and squarks the primary quark
jets have so little energy they would not pass the cut for an isolated
jet so would present themselves in much the same way as simple $S_q
S_q^*$ production. 

Now let us turn to the issue of deciding, given a putative excess of
events with $n_{jets} \ge 4$, whether the excess can be attributed to 
production of $S_q S_q^*$.  Fortunately for the discussion here, some
important characteristics of the final states originating from $S_q
S_q^*$ are the same as for the SUSY-Higgs search which was the
motivation for the ALEPH analysis. Therefore many of the quantities
needed in order to decide whether events are consistent with
coming from $S_q S_q^*$ pairs have already been calculated and
reported by ALEPH$.^5$ 

A very important characteristic of squark pair production is that
the squarks produced are degenerate in mass.  This is because gauge
interactions (including their SUSY-transforms involving gauginos) 
conserve chirality.  Moreover the absence of flavor-changing
neutral currents implies that gauge interactions of squarks are
flavor-diagonal to high accuracy.  Thus when a squark and antisquark
pair is produced in $e^+ e^-$ or hadron colliders, their flavor and
chirality are the same.  Furthermore, the {\it mixing} between
eigenstates of chirality for a given flavor squark is small in this
scenario, except for the stops\footnote{When the effective theory
contains only dimension-2 SUSY breaking operators, $A=0$.  Then the
mixing $\delta_q$ between chirality eigenstates for squark flavor $q$
is $\delta_q = \frac{\mu m_q}{{M_q}^2}$ times $cot \beta$ ($tan
\beta$) for charge 2/3 (-1/3) squarks respectively; ${M_q}$ is the
average mass of the squarks of flavor $q$. }.  Thus except for stop
pairs, {\it the correct pairing of jets from decays of a squark pair
will produce equal mass dijets}.  This is a crucial point.  Since the 
various squark flavors need not be degenerate, the dijet invariant
mass spectrum may be messy, with nearby overlapping peaks or
enhancements.  Nonetheless, a clear signal is possible since correct
pairing of jets always leads to a vanishing {\it difference} of the
dijet invariant masses.  Henceforth jets are always taken to be paired
so the dijet mass difference is minimized.

Out of the 14 ALEPH events, 8 have a total dijet mass centered on 109
GeV, with a spread $\sim 10$ GeV consistent with resolution.  All
eight of these events have a dijet mass difference compatible with
zero, as will be discussed in the next paragraph.  Thus these 8 events
are candidates for originating from pairs of $\sim$ 54.5 GeV squarks.
This is exactly the squark mass required to account for the observed
number of excess events, for the dls case illustrated in Fig. 1.

If it is correct to interpret an excess of events with $n_{jet} \ge 4$
as due to production of approximately degenerate $S_q S_q^*$'s, the
excess should be a definite function of cm energy.  The $r_i$'s of eq.
(1) grow as $[1-( 2 M_{\rm dijet} /E_{cm})^2)^{\frac{3}{2}}$.
Ignoring the variation of $f_{\ge 4}$ with energy, this means that if
the ALEPH excess were due to production of degenerate $S_q S_q^*$
pairs, when LEPII runs at 165 (190) GeV the $\ge 4$-jet event rate
should be 3.1 (3.7) times the expected rate from standard model
processes alone.  Thus with improved statistics and higher energy,
LEPII measurement of $R_{\ge 4}$ will provide a powerful tool to
support or exclude the hypothesis that squarks are being produced,
which then decay to quarks and hadronizing-gluinos.   

What more can be done with a given event sample?  Unfortunately,
the prediction that the squark and antisquark are mass degenerate on
an event by event basis may not provide a useful identifier for 
squarks at present energies. For instance, at $E_{cm} \sim 135$ GeV
ALEPH's resolution in dijet invariant mass is about 20 GeV
full-width-at-half-max.  Although the distribution in
dijet-mass-difference of the 14 ALEPH events is consistent with equal
dijet invariant masses, requiring the minimum dijet mass difference to
be less than 20 GeV only reduces the number of events expected in the
standard model from 8.6 to 7.1$.^5$  This implies that at $E_{cm} =
135$ GeV, 80\% of standard model events have a dijet mass difference
less than 20 GeV, when jets are paired so as to minimize the dijet
mass difference and their other cuts are satisfied.  Hence the dijet
invariant mass difference does not at this energy and squark mass
provide a useful test as to whether the excess events are $S_q S_q^*$
in origin, let alone provide a discriminant as to which $\sim 7$
events are potentially squarks and which $\sim 7$ are ordinary (mainly
$q \bar{q} g g$) events.     

However there are other discriminants which can be investigated.
Spin-0 particles produced in $e^+ e^-$ scattering through a spin one
photon or $Z^0$ have a $sin^2 \theta$ angular distribution.  After
determining the dijet 3-momenta, the angular distribution of the
dijets can be formed.  If the events with total dijet mass $\sim 109$
GeV were due to the decay of squark pairs, taking these events alone
should produce an angular distribution $\sim sin^2 \theta$.  The
remaining events (presumably comprised of $q \bar{q} g g$) should be
produced according to the standard model and thus have a different
characteristic angular dependence.  Hopefully the two distributions
will prove qualitatively different enough to allow true $S_q S_q^*$ to
be distinguished from background. If the situation is more
complicated, as in the cocktail mentioned above with charginos and
squarks, large statistics will probably be necessary to compare
expected and predicted distributions. 

Still more detailed investigations of the events can be made.  Some of
the salient characteristics to explore are:

I.  Certain correlations in the jet angular distributions should reflect
the fact that the dijets are actually spin-0 particles, which decay to
spin-1/2 particles which then form jets.

II. $S_q S_q^*$ events have two gluino jets, one associated with each
dijet. On account of their larger color charge, gluino jets may be
``fatter'' than standard quark jets.  The hadronization of gluino jets
will nearly always produce an $R^0$\cite{f:95}.  Denote the average
momentum fraction of an $R^0$ with respect to its jet by $x_R$.  $x_R$
could be determined in a Monte Carlo or other model of jet 
fragmentation, or taken by analogy from, say, charm fragmentation.  The
$R^0$'s will decay to a photino and a small number of
pions\cite{f:102}.  The photino will typically have a momentum
transverse to the jet axis of $\sim 0.4 - 0.8$ GeV\cite{f:102},
depending on the relative mass of $R^0$ and $\tilde{\gamma}$.  If the
lifetime of the $R^0$ is short compared to the transit time of the
calorimeter and its decay is two-body, the photino will
typically have a momentum along the jet of $ \lsi x_R \frac{
M_{sq}}{2}$.  This will lead to characteristic small deviations from
the energy and momentum accounting in the event which will show up in
appropriate variables. If on the other hand the $R^0$ lifetime is long
enough that it loses its kinetic energy in the calorimeter before
decaying, the momentum along the jet axis carried away by the photino
will be nearly imperceptible.  A final possibility for
``$R^0$-tagging'' is to look for $\eta$'s, which should be produced
with branching fraction $\sim 0.1$ if $R^0$ decay is predominantly
2-body\cite{f:99}. Resolution might not be adequate for this, even if
enough $\ge $4-jet events were available.  

III.  The dominant squark decay channel, $S_q \rightarrow q \tilde{g}$,
can produce two jets or bremstrahlung additional gluons forming $\ge
2$ jets, just as QCD produces $\ge 2$-jet final states in $Z^0
\rightarrow q \bar{q} $ decay.  Indeed, a non-negligible fraction of
squark final states can be expected to contain $\ge 3$ jets, as in
$Z^0$ decay.  With typical jet definitions (e.g., $y_{cut} = 0.01$),
40\% of the hadronic $Z^0$ decays have $\ge 3$ jet final states.
The larger color charge of the gluino is likely to enhance
bremstrahlung compared to a quark jet, although when the squark is
significantly lighter than the $Z^0$ this is at least partially
compensated by the reduction in phase space which causes jets to
coalesce more.  These issues should be straightforward to model with
minor modifications to event generators such as Pythia and Herwig.   

IV.  At 133 GeV the production rates for squarks of different flavors
is the same, aside from the $\beta^3$ factor which differs if they are
not mass degenerate.  If the ALEPH excess of events persists when
statistics improve and the excess is associated with a single 
peak in invariant mass, as is consistent with the present small
sample, there should be approximately equal numbers of events for each
squark flavor produced.  When the sbottoms and stops are below
threshold, the only $b \bar{b}$ events would be due to standard model
final states and $b \bar{b} b \bar{b}$ would be extremely rare. 

To summarize: I discussed the events with four or more jets which
result from $S_q S_q^*$ production followed by decay to quark and
gluino.   When SUSY breaking does not produce tree-level gaugino
masses or scalar trilinear couplings, the gluino is light and
hadronizes so such events are the dominant end product of squark pair
production.  Methods were given to decide whether an observed 4-jet
excess is due to this process.  If the excess of 4-jet events reported
recently by ALEPH is confirmed by other experiments and higher
statistics, it could be circumstantial evidence that some flavors of
squarks have mass around 55 GeV.  These squarks would be decaying into
a quark and gluino which hadronize to two or more jets.  This
hypothesis is consistent with several features observed in the present
small sample:   
\begin{enumerate}
\item  The number of excess 4-jet events, 14 observed when 7.1 were
expected, is the number expected with 5 flavors of $L$-squarks having
masses $\sim 55$ GeV.  In fact, 8 of the 4-jet events do have a total
dijet mass of 109 GeV.  (But note that other arguments make it unlikely
for five squarks to be this light.)
\item  The observed peaking at zero dijet mass difference is expected
because of the mass-degeneracy of the $S_q S_q^*$ pair for each
flavor.  (But note that the dijet mass resolution is such that this is
not a very stringent test.)
\item  The amount of missing energy in each of the events is
small -- this is as expected because the photinos which are lost
are produced after hadronization of the gluino, so they carry very
little energy.  
\item  No $b \bar{b} b \bar{b}$ events are seen; they are not expected
in the present scenario because two of the final jets must be gluino
jets.  
\end{enumerate}
Whether ALEPH has seen a first hint of the production of squarks or
squarks and charginos can be investigated further with the
existing data.  It will be readily confirmed or refuted by more
statistics and higher energies.  In any case, careful study of 4-jet
events should be a standard part of squark search techniques until the
time that a light, hadronizing gluino has been excluded.  

{\bf Acknowledgements:}  I have benefited from discussing these matters
with L. Clavelli, J. Conway, S. Lammel and R. Rattazzi.




\end{document}